\documentclass[12pt]{article}
\usepackage{graphicx,psfrag,epsf}
\usepackage{enumerate}
\usepackage[round]{natbib}
\usepackage{url} 
\usepackage{amsmath,amssymb,amsfonts,amsthm}
\usepackage{float,graphicx, multirow, setspace}
\usepackage{subfiles}
\usepackage{placeins}
\usepackage{url}
\usepackage{hyperref}
\usepackage{xr, xr-hyper}
\usepackage{algpseudocode}
\usepackage{subfig}
\usepackage{algorithm}
\usepackage{makecell}
\usepackage[scr=boondox]{mathalfa}

\usepackage{afterpage}

 \renewcommand{\P}{ \mathbb{P} }

 \newcommand{\xCTRL}{{ \mathrm{xCTRL} }}

 \newcommand{\bx}{{ \mathbf{x}}}

 \newcommand{\fixme}[1]{\textcolor{orange}{\small [fixme]: #1}}

 \usepackage[normalem]{ulem}
\usepackage[dvipsnames]{xcolor}

\newcommand{\blind}{0}

\begin{document}

\bibliographystyle{apalike}

\def\spacingset#1{\renewcommand{\baselinestretch}%
{#1}\small\normalsize} \spacingset{1}


\if0\blind
{
  \title{\bf{Separating Intent from Execution: A Probabilistic Approach to Pitch Location Accuracy}}
  \author{
    Matt Ludwig\thanks{Correspondence to: mludwig239@gmail.com}, \
    Ryan S. Brill\thanks{
      Graduate Group in Applied Mathematics and Computational Science, University of Pennsylvania
    }, \ 
    and Abraham J. Wyner\thanks{
      Dept.~of Statistics and Data Science, The Wharton School, University of Pennsylvania
    }
  }
  \maketitle
} \fi

\if1\blind
{
  \bigskip
  \bigskip
  \bigskip
  \begin{center}
  \fixme{}
  \end{center}
  \medskip
} \fi

\bigskip
\begin{abstract}
Control has long been recognized as a critical component of pitcher performance, reflecting a pitcher's ability to execute pitches in alignment with his intended targets. However, accurately inferring a pitcher's intentions presents a persistent challenge. Traditional metrics typically rely on uniformity assumptions, inferring intent based on the behavior of a ``typical'' pitcher across similar situations. In this study, we propose an alternative, individualized approach to measuring control, one that eschews such assumptions in favor of personalized inference. We estimate a pitcher's intended location on a pitch-by-pitch basis, conditioning on both individual tendencies and specific game contexts. This allows us to assess control by comparing the actual pitch location to the inferred intended target, thereby aligning measurement more closely with the unique strategies of each pitcher. We introduce xCTRL, a novel metric that quantifies control as the distance between a pitch's actual location and its estimated intended location. We find that xCTRL exhibits strong stability and greater predictive power than existing control metrics. By capturing pitcher-specific intent, xCTRL enhances our understanding of control and offers a more intuitive and accurate representation of pitching performance.
\end{abstract}

\noindent%
\vfill

\newpage
\spacingset{1.45} 

\section{Introduction}\label{sec:intro}

What makes a pitcher successful? While this question may seem straightforward, it is quite difficult to answer in practice. Analyses typically break down a pitcher's effectiveness into two main components: \textit{stuff} and \textit{control}. At a high level, stuff refers to how difficult a pitch is to hit––capturing factors like velocity, movement, and release point––while control reflects a pitcher's ability to locate pitches where he intends to. In recent years, the baseball analytics community has made significant progress in evaluating stuff, but control has proven much harder to quantify.
We believe this gap stems from a fundamental challenge: it is impossible to directly observe where a pitcher was trying to throw a given pitch. 

Existing control metrics compare outcomes to the behavior of a hypothetical ``optimal'' pitcher. These approaches rely on uniformity assumptions, treating all pitchers as though they share the same intent and target similar zones. They assume every pitcher tries to follow the same ''optimal'' policy. But this is a flawed premise. Control should reflect a pitcher's ability to execute on \textit{his own} intentions, not some league-wide standard.

To address this, we introduce a new control metric––\textit{xCTRL}––that measures execution relative to individualized intent. Although a pitcher's intent is not directly observable, we can infer it statistically by combining the actual pitch location with the pitcher's historical location data. With sufficient data, we can estimate where a pitcher tends to locate each of his pitch types in a given context. If a pitch lands near one of these typical targets, we interpret it as evidence of good execution.

Unlike traditional metrics, xCTRL does not rely on league-wide expectations of where a pitcher \textit{should} target. Instead, it assesses control based solely on the pitcher's own location tendencies, removing the need for assumptions of pitcher uniformity. Visualizations in Section~\ref{subsec:viz} illustrate the diversity in location strategies, even among top pitchers. By modeling each pitcher's unique targeting behavior, xCTRL provides a fair and individualized measure of control.

xCTRL is both interpretable and practical. It has strong predictive value for key pitcher performance metrics, remains stable across seasons, and aligns with intuitive expectations. Pitchers with consistent success tend to score better in xCTRL, while those with erratic results tend to score worse. By quantifying control as the ability to execute on personal intent, xCTRL offers a significant improvement over previous approaches and complements modern evaluations of stuff. It enables more nuanced and accurate assessments of player value.

\subsection{Traditional Measures of Control}\label{sec:history}

Early measures of control were based on simple at-bat outcomes. Walks, in particular, have long been viewed as indicative of poor control. Accordingly, metrics such as WHIP (Walks and Hits per Inning Pitched) and BB/9 (walks per nine innings) have been used to assess control.

These outcome-based metrics are coarse and limited. They operate at the at-bat level and cannot capture pitch-level nuance, especially variation in control across pitch types within the same at-bat. A more refined approach uses the location of individual pitches, which is a natural choice since control is inherently a pitch-level concept.

Pitch location data (i.e., recording where the ball crosses the strike-zone plane) has been publicly available since the 2008 MLB season,\footnote{
    \url{https://baseballsavant.mlb.com/csv-docs}
} enabling more granular control metrics. The most widely used location-based measure is Location+ \citep{stuffprimer}, which evaluates control by comparing pitch outcomes to league-wide expectations for each count and pitch type. However, Location+ assumes that all pitchers should be targeting the same ``optimal'' zones, conflating control with conformity and failing to account for individual strategy.

Others have recognized the need to evaluate control relative to a pitcher's own tendencies \citep{command}. Command+, for example, models pitcher intent as a discrete selection from a finite set of targets and incorporates contextual cues—such as catcher glove placement—to estimate intention. While this is a step forward, the method has important limitations: reliance on discretized targets can limit precision, and its complexity may restrict broader adoption.

Critically, existing metrics like Location+ and Command+ lack a key ingredient: {a posteriori} correction. They predict where a pitcher is likely to aim before the pitch is thrown, but do not update those beliefs using the pitch’s actual location. As a result, they fall short of inferring the intent behind each specific pitch––an essential step for accurately measuring control.

\subsection{Limitations of Traditional Measures of Control}\label{sec:issues}

Traditional measures of pitcher control suffer from three major limitations, each of which is addressed by xCTRL.

\textbf{Problem 1: Reliance on at-bat outcomes.}  
Measuring control based solely on at-bat outcomes (e.g., WHIP, BB/9, and K/BB) ignores most of the pitches thrown. Many at-bat outcomes are based on the final pitch, overlooking all other pitches. For instance, a pitcher may dot the corners with multiple pitches but allow a hit on one poorly located pitch. Outcome-based metrics would penalize him for poor control, despite the majority of his pitches being well-executed.

Such metrics also assume that all walks are uniformly negative, which oversimplifies the strategic nature of pitching. In reality, a walk may reflect intentional and well-controlled pitching in a specific context. If pitch locations align with a pitcher's typical targeting strategy, the walk may have been deliberate. Metrics based purely on outcomes conflate intention with failure. xCTRL distinguishes between control failures and strategic decisions by grounding inference in pitch-level location data.

\textbf{Problem 2: Assumptions of pitcher uniformity.}  
Location-based metrics such as Location+ assume all pitchers share common location goals for each pitch type and count. This imposes a standard of league-wide optimal targets and penalizes pitchers who deviate, even when those deviations reflect a deliberate and repeatable strategy. As a result, pitchers with excellent control but unconventional location policies may be undervalued. Visual examples in Section~\ref{subsec:viz} illustrate this flaw clearly.

Command+ takes a step toward accounting for individuality, but still imposes a limited and discrete set of possible target zones––thirteen, each tied to a traditional pitch type and location pairing. This framework presumes all pitchers conceptualize location identically, preventing fine-grained representation of their preferences. For example, two pitchers both classified as aiming ``up-and-in'' may in fact target meaningfully different regions, but Command+ lacks the resolution to capture this. The result is systematic mischaracterization of intent. xCTRL avoids this issue by modeling each pitcher’s intent continuously and independently.

\textbf{Problem 3: Disregarding the realized pitch location.}  
Finally, existing control metrics do not incorporate the actual pitch location when inferring intent. Both Location+ and Command+ predict where a pitcher is likely to aim, then treat that prediction as ground truth intent. But the observed pitch location provides valuable information about the pitcher’s likely target. For example, if a pitcher is equally likely to target two zones, and the pitch lands near one of them, this observation should shift our belief toward that target. Failing to update based on the realized outcome leaves inference incomplete.

A robust control metric should use both a prior over likely intent and a posterior adjustment incorporating the pitch’s final location. xCTRL does exactly this: it estimates a pitcher’s intent using historical tendencies, then refines that estimate using the actual pitch location. This yields a more accurate and individualized measure of control.

\subsection{Devising a Better Measure of Control}

Motivated by the need to account for pitcher individuality and to improve intent estimation, we introduce a new framework for measuring control that avoids the limitations of traditional metrics. The main contributions of xCTRL are twofold: (1) identifying individualized pitcher tendencies and (2) incorporating post-pitch adjustments to refine intent inference. Specifically, we use historical pitch location data to fit a multimodal distribution of targeting tendencies for each combination of pitcher, pitch type, and batter handedness. These distributions are estimated using the EM algorithm to fit a Gaussian Mixture Model (GMM), which represents the pitcher’s targeting behavior in a given game-state. The observed pitch location is then used to compute the posterior probability that the pitcher was aiming at each of these typical targets. This enables xCTRL to separate intent from execution error on a pitch-by-pitch basis.

By using every pitch rather than just at-bat outcomes, xCTRL eliminates wasted information and avoids the conflating factors inherent in outcome-based metrics. Like Location+, xCTRL operates at the pitch level, allowing for analysis by pitch type. However, xCTRL is more robust because it establishes location expectations separately for each pitcher, avoiding the assumption of a single league-wide optimal strategy. 
As a result, our metric does not conflate control with a pitcher's conformity to league norms. 
Further, xCTRL does not assume that a pitcher's location is solely determined by context. It uses a posteriori adjustment––conditioning on the actual pitch location––to infer the most likely target for each specific pitch.

xCTRL also improves upon Command+ by modeling the strike zone as a continuous space in $\mathbb{R}^2$, rather than relying on manually defined discrete partitions. This flexibility allows pitchers to be evaluated fairly regardless of where they tend to aim, eliminating bias that might favor those who happen to target predefined zones. In doing so, xCTRL avoids the rigid assumptions that underlie existing control metrics like Command+.

Despite these advantages, xCTRL has a notable limitation: it requires sufficient data to accurately estimate targeting distributions. Tendency models are constructed for each pitcher, pitch type, and batter handedness combination, and reliable inference depends on having a reasonably large number of observations. In low-data settings—such as when conditioning on count—this requirement can limit the scope of analysis. To address this, we propose a method for extending xCTRL to low-data contexts, detailed in Appendix~\ref{app:weighted_em_count}.

The remainder of this paper is organized as follows. Section~\ref{sec:DataAndModel} outlines the details of our Gaussian mixture modeling approach, which we fit using the EM algorithm. In Section~\ref{sec:results}, we present the results of our analysis and compare xCTRL to existing control metrics. Finally, in Section~\ref{sec:discussion} we discuss potential applications of xCTRL and directions for future improvement.

\section{Data and model specification}\label{sec:DataAndModel}

We begin in Section~\ref{sec:data} with a brief overview of our MLB pitch-by-pitch dataset, highlighting key variables for measuring pitch control. In Section~\ref{sec:model}, we introduce our Gaussian mixture pitch location density model that forms the basis of our definition of control. In Section~\ref{sec:em} we estimate the parameters of the model and in Section~\ref{sec:boot} we quantify uncertainty in these estimates.

\subsection{Data}\label{sec:data}

We calculate xCTRL entirely from pitch location data. We compiled our dataset by scraping Statcast using the \texttt{pybaseball} package \citep{pybaseball}, covering every regular season pitch from 2008 to 2023. Each pitch is binned by pitcher, batter handedness, and pitch type (e.g., fastball, curveball, etc.), though Statcast allows for more granular binning (e.g., by count) if desired. Unless stated otherwise, all xCTRL results in this paper use bins defined by pitcher, batter handedness, and pitch type. To account for year-to-year changes in pitch location tendencies (e.g., see Figure~\ref{fig:verlander_heatmap}), we compute xCTRL separately for each season.

Pitch location refers to the $(x, z)$ coordinates where the pitch crosses the strike-zone plane. The $x$-axis runs horizontally, with $x = 0$ at the center of home plate; negative $x$ values lie to the catcher's left. The $z$-axis represents height above the ground. Statcast reports location data in feet, but we report xCTRL in inches to provide a more intuitive sense of control error.


Part of xCTRL's strength lies in its ability to predict pitcher success. To demonstrate this, we collect season-level performance metrics such as Fielding Independent Pitching (FIP), Innings Pitched, and BB/9, showing that xCTRL is a stable and effective forecasting tool. We also analyze pitch-level outcomes––such as RE24 and the change in expected runs––to highlight the granularity of xCTRL that is not possible with Location+.

While control is crucial, it is not the only determinant of pitching success. Pitch quality, particularly in the modern game, plays a major role. To capture pitch quality, we use FanGraphs' Stuff+ metric \citep{stuffprimer}, which fuses variables such as velocity, movement, and release point to estimate how difficult it is to hit a particular pitch. Stuff+, however, is proprietary and not available at the pitch level.
We also scrape Location+, FanGraphs' control metric discussed in Section~\ref{sec:intro}, to compare with xCTRL. Stuff+ and Location+ were introduced in 2020, but due to limited data in the COVID-shortened season, we exclude 2020 from our analysis. Altogether, we have aggregated xCTRL, Stuff+, and Location+ for each pitch type thrown by a pitcher during the 2021-2023 MLB seasons.

\subsection{Calculating xCTRL via a Gaussian Mixture Model}\label{sec:model}

Simply put, xCTRL measures the distance between the true location of a pitch and its inferred intention. Consider a pitch that crosses the plate at location $(x_\ast, z_\ast)$.
If the pitcher's intended target was $(x', z')$, a natural measure of control on that pitch is the Euclidean distance between the actual and intended locations:
\begin{equation}
    \delta((x_\ast, z_\ast), (x', z')) := \sqrt{ (x_\ast - x')^2 + (z_\ast - z')^2 }.
\label{eqn:delta}
\end{equation}

As it is impossible to peer inside the mind of a pitcher, we do not know the true intended pitch location $(x', z')$. The main accomplishment of xCTRL is an enhanced method for inferring $(x', z')$. xCTRL estimates this using pitcher tendencies and post-pitch likelihood adjustment. Taking a probabilistic approach, xCTRL begins with a parameterized distribution representing pitcher location tendencies. This distribution represents initial beliefs about where the pitcher aims, before adjusting for the realized location.

We model the density of pitcher tendencies using a Gaussian mixture model (GMM). Each component of the mixture represents a unique target the pitcher likes to aim at. The component weights reflect how often each target is thrown to. Each distribution contains $K$ components, where $K$ is a tunable hyperparameter that allows xCTRL to flexibly capture varying levels of complexity in a pitcher's location strategy. We fit a separate GMM for every combination of covariates $\bx$, defined by pitcher, pitch type, batter handedness, and season.

The distribution representing a pitcher's behavior has interpretable parameters. It consists of $K$ components, each corresponding to a distinct intended target. They are not targeted uniformly: the weight of each component, $\pi_k(\bx)$, reflects the frequency with which the $k^{\text{th}}$ target is selected in the context defined by bin $\bx$.
Each component has a mean $(\mu^x_k(\bx), \mu^z_k(\bx))$, representing the center of the target. 
However, we understand that pitchers are not aiming at one incredibly precise point––they are typically trying to induce some desirable outcome (e.g., jamming the batter with an inside fastball). 
To capture this, each component has an associated covariance matrix, $\Sigma_k(\bx)$, which governs the shape and orientation of the target region. This allows for some targets to be tightly clustered, while others may be broader or elongated in specific directions to reflect side-to-side or vertical variation.
The Gaussian mixture model offers both flexibility and precision for capturing individual pitcher tendencies. 
We can represent a variable number of targets per pitcher, their locations, how tight they are, and their orientation, all conditioned on the covariates $\bx$.

The Gaussian mixture model provides a reliable representation of a pitcher's general pitch location policy. It encodes prior beliefs about where a pitcher is likely to aim before the pitch is thrown. To infer intent on a pitch-by-pitch basis, xCTRL applies a posterior adjustment using the observed pitch location. Specifically, the actual location is used to update the likelihood that the pitcher was aiming at each of the $K$ target zones identified by the mixture model. This posterior update improves upon relying solely on the prior component weights and allows xCTRL to estimate intent for each individual pitch, rather than only at the bin level $\bx$.

While we cannot know which target the pitcher intended, we can infer the probability that he aimed at each of his $K$ established targets. 
In particular, we compute the posterior probability that he aimed at each target given the observed pitch location:
\begin{align}
  p_k((x_\ast, z_\ast), \bx) 
  &:= \P(\text{aim at cluster } k \mid (x_\ast, z_\ast), \bx) \\
  &= \frac{ \P((x_\ast, z_\ast) \mid \text{aim at cluster } k, \bx) \cdot \P(\text{aim at cluster } k \mid \bx) }{\P((x_\ast, z_\ast) \mid \bx)} \\
  &= \frac{ \P((x_\ast, z_\ast) \mid \text{aim at cluster } k, \bx) \cdot \P(\text{aim at cluster } k \mid \bx) }{ \sum_{j=1}^{K} \P((x_\ast, z_\ast) \mid \text{aim at cluster } j, \bx) \cdot \P(\text{aim at cluster } j \mid \bx) } \\
  &= \frac{ \mathcal{N}((x_\ast, z_\ast); \mu_k(\bx), \Sigma_k(\bx)) \cdot \pi_k(\bx) }{ \sum_{j=1}^{K} \mathcal{N}((x_\ast, z_\ast); \mu_j(\bx), \Sigma_j(\bx)) \cdot \pi_j(\bx) }.
\end{align}
Here, $\mathcal{N}$ denotes the bivariate Gaussian density. 

If the pitcher was aiming at cluster $k$, how well did he execute his intent? We define execution as the Euclidean distance between the observed pitch location and the center of cluster $k$. Specifically, if the intended target was cluster $k$, his execution is
\begin{equation}
    \delta_k((x_\ast, z_\ast)|\bx) := \delta\big( (x_\ast, z_\ast), (\mu^x_k(\bx), \mu^z_k(\bx)) \big).
\label{eqn:delta_k}
\end{equation}
Although the target centers are the only mixture parameters directly used to measure execution, the shape of each target remains essential. It plays a critical role in the posterior update that determines the proability vector over intended targets.

We then define xCTRL for an individual pitch by the dot product between the posterior probability vector and the vector of execution distances.
In other words, it is a probability-weighted sum of execution values:
\begin{equation}
    \xCTRL((x_\ast, z_\ast), \bx) := \sum_{k=1}^{K} p_k((x_\ast, z_\ast), \bx) \cdot \delta_k((x_\ast, z_\ast)| \bx).
\label{eqn:control}
\end{equation}
A pitcher's overall xCTRL for bin $\bx$ is defined by the average xCTRL across all pitches in that bin.

Since xCTRL is a weighted distance metric expressed in inches, lower values indicate better control. Specifically, xCTRL quantifies the average deviation between a pitcher's observed pitch locations and their inferred intended targets. We provide a visualization of this calculation in Section~\ref{subsec:viz}.

\subsection{Fitting the mixture using the EM Algorithm}\label{sec:em}

xCTRL measures a pitcher's control conditional on covariates $\bx$ (e.g., Gerrit Cole's fastball against left-handed batters in 2023). The first step is to model the pitcher's location tendencies in this situation. We use a Gaussian mixture model. We fit the parameters of this model using the Expectation-Maximization (EM) algorithm \citep{originalem}, implemented by the \texttt{scikit-learn} library in Python \citep{gmsk}. Fitting a Gaussian mixture in continuous space allows xCTRL to operate with greater accuracy than methods that discretize the strike-zone plane.

The EM algorithm requires specification of the number of mixture components. To capture pitcher individuality, xCTRL does not assume a fixed number of components. Instead, the number of components is a function of the covariates: $K = K(\bx)$. We treat $K$ as a tunable hyperparameter. For each bin $\bx$, we randomly partition the dataset into training and validation sets, fit models for various values of $K$ (typically $K \in \{1, \dots, 6\}$) on the training data, and select the value of $K$ that maximizes the log-likelihood of the validation set.

Before fitting the Gaussian mixture model via the EM algorithm, location outliers are removed to ensure more reliable parameter estimation. Outliers are identified using the 1.5 interquartile range (IQR) rule: any pitch with an $x$- or $z$-value more than $1.5 \cdot \text{IQR}$ from the mean along either dimension is excluded during the fitting process. However, xCTRL scores are still computed for these outliers after the mixture has been fit.

xCTRL requires a fairly large number of pitches to stabilize. We set a minimum threshold of 250 pitches for fitting a tendency distribution within a bin $\bx$. This ensures the EM algorithm has sufficient data to distinguish among multiple meaningful targets. For this reason, our analysis restricts covariates $\bx$ to pitcher, pitch type, and batter handedness within each season. xCTRL can be extended to finer-grained contexts when data volume permits. We discuss an extension for low-data settings in Appendix~\ref{app:weighted_em_count}.

It is also important to determine which location data should be used to fit a pitcher's tendency distribution. Because a pitcher's location policy may evolve over time, location data must be partitioned to reflect consistent policy periods. Otherwise, the EM algorithm may fit a single distribution to data reflecting multiple distinct strategies, resulting in an unreliable control estimate. Pitcher policies may change for many reasons (e.g., mechanical adjustments, injury recovery, or coaching changes). As a general rule, we treat each season as the potential beginning of a new policy. This design choice improves xCTRL’s accuracy by avoiding policy confounding, though it limits the number of pitchers we can evaluate due to smaller seasonal sample sizes.

\subsection{Quantifying Uncertainty}\label{sec:boot}

xCTRL measures execution relative to inferred targets, but because a pitcher's true intent is never observed, it is important to quantify the uncertainty in these estimates. The first step in computing xCTRL is fitting a pitcher’s tendency distribution using the EM algorithm. However, the EM algorithm is not deterministic––random initializations and variations in training/validation splits can lead to different Gaussian mixture parameterizations across runs––providing another source of uncertainty.

To address this, we use bootstrapping to quantify uncertainty. For each bin $\bx$, we generate 100 bootstrapped resamples of the associated pitch locations using the standard i.i.d.\ bootstrap, which resamples pitches uniformly with replacement. Each resample is used to fit a new Gaussian mixture model and compute the corresponding xCTRL scores.
This results in 100 xCTRL values per bin, from which we compute a $90\%$ confidence interval using the $5^{\text{th}}$ and $95^{\text{th}}$ percentiles. All xCTRL values reported in this paper are the median from the 100 bootstrap samples. Likewise, all visualizations reflect the fitted Gaussian mixture associated with the median xCTRL estimate.

\section{Results}\label{sec:results}

xCTRL is computed solely from pitch location data. Statcast has publicly provided this data since the 2008 MLB season. Most of our analysis focuses on the 2021--2023 seasons, due to limitations in the auxiliary variables considered alongside xCTRL. Because xCTRL relies only on location data and requires no proprietary inputs or complex models, it is easily extensible, enabling retroactive or more detailed analyses by anyone interested.

Our primary focus is on fastballs, as control for fastballs is widely considered more important than for other pitch types. That said, we also demonstrate that xCTRL is predictive of success across other pitch types, even though they are not the main focus of this paper.

This section proceeds as follows. In Section~\ref{subsec:viz}, we visualize fitted Gaussian mixtures representing pitcher location tendencies. In Section~\ref{sec:control_ranking} we compare xCTRL to Location+. In Section~\ref{sec:model_validation} we show that xCTRL is more predictive than existing metrics of key indicators of pitcher success. Finally, in Section~\ref{sec:moreAnalyses} we highlight additional analyses enabled by xCTRL that were previously infeasible with traditional metrics.

\subsection{Visualizing pitch location distributions}\label{subsec:viz}

The first step in computing xCTRL is estimating a pitcher's location tendencies. We model this using a Gaussian mixture fitted via the EM algorithm. The mixture parameters represent the number, center, and shape of a pitcher's typical targets. Unless otherwise stated, the visuals in this section depict four-seam fastballs thrown to right-handed batters during the 2023 MLB season. We visualize pitch location distributions using heatmaps generated by the \texttt{seaborn} library in Python \citep{seaborn}, representing the strike zone by a red rectangle based on average dimensions from \citet{strikezone}. All visuals are presented from the catcher's perspective, consistent with Statcast coordinates.

\begin{figure}[hbt!]
\centering
\includegraphics[width=0.5\textwidth]{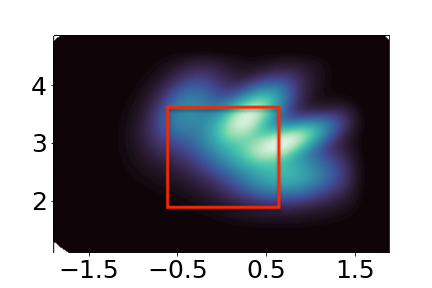}
\caption{
  Gerrit Cole's 2023 pitch location density for four-seam fastballs against right-handed batters.
}
\label{fig:cole_heatmap}
\end{figure}

We begin in Figure~\ref{fig:cole_heatmap} with Gerrit Cole, who won his first Cy Young Award in 2023 and is widely known for the quality of his fastball. Cole's heatmap reveals four distinct fastball targets clustered along the upper and outer edges of the strike zone. The most frequent target is up-and-away, while up-and-in is used least often. These patterns represent Cole's \textit{overall} tendencies when throwing fastballs to right-handed hitters. Of course, his targeting choices may vary based on other factors. For instance, he may be more likely to go up-and-in with two strikes while seeking a strikeout. Without sufficient data, it’s difficult to infer how these tendencies shift in more specific game states. This is precisely where xCTRL’s post-pitch intent adjustment becomes useful––it refines our understanding of intent for each pitch using both historical tendencies and the realized pitch location.

\begin{figure}[htb!]
    \centering
    \subfloat[]{{\includegraphics[height=0.25\textwidth]{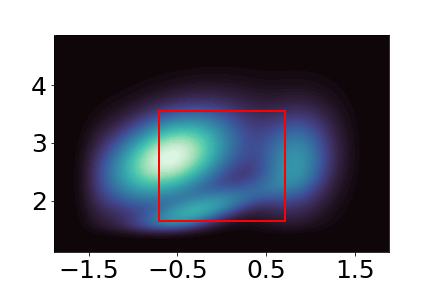}} \label{fig:friedrhb} }
    \qquad
    \subfloat[]{{\includegraphics[height=0.25\textwidth]{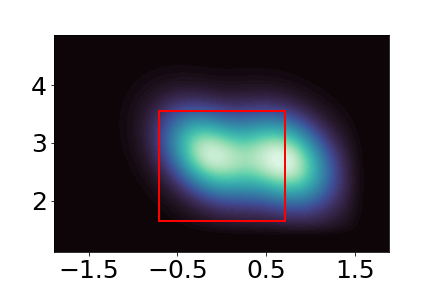}} \label{fig:wheelerrhb}}
    \qquad
    \subfloat[]{{\includegraphics[height=0.25\textwidth]{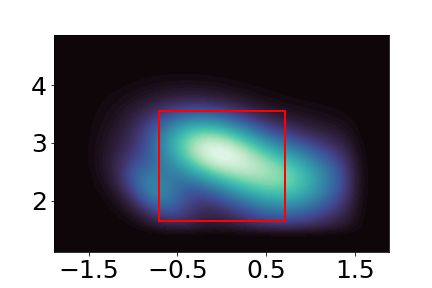}} \label{fig:scherzerrhb}}
    \qquad
    \subfloat[]{{\includegraphics[height=0.25\textwidth]{ColeRHBFF2023.png}} \label{fig:colerhb}}
    \qquad
\caption{
  2023 fastball intent distributions against right-handed batters for Max Fried (a), Zack Wheeler (b), Max Scherzer (c), and Gerrit Cole (d).
}
    \label{fig:top_4_starters}
\end{figure}

Gerrit Cole has a clearly defined location policy, but what about other starting pitchers? The uniformity assumption underlying Location+ implies that all successful-control pitchers should target the same zones. We assess this assumption using pitch tendency visualizations in Figure~\ref{fig:top_4_starters}, which show the fastball location strategies of several of the most successful starting pitchers in recent seasons (as identified by \citet{bestStarters}).

Unsurprisingly, the results offer strong evidence against pitcher uniformity. None of these elite pitchers follow the same location strategy. Gerrit Cole and Max Fried both work the edges of the strike zone, but in distinctly different ways. Zack Wheeler and Max Scherzer appear most similar at first glance, yet meaningful differences emerge upon closer inspection. Wheeler primarily targets laterally at the waist, while Scherzer spans a diagonal spectrum from up-and-in to low-and-away. Scherzer also features an infrequent low-and-inside target, possibly to remain unpredictable.

It is unlikely that these pitchers could sustain elite performance using suboptimal location strategies. A more plausible explanation is that no universal location policy exists. Each pitcher succeeds by tailoring his approach to his own strengths. It then follows naturally that control should be measured relative to individual intent, not against a league-wide benchmark.

It is important to specify the time frame over which location tendencies are estimated. It would be unreasonable to assume that a pitcher follows a fixed location strategy throughout his entire career. Pitchers may adjust their approach for many reasons––adding a new pitch, working with a different catcher, or simply trying to remain unpredictable. To avoid conflating distinct strategies, we fit pitcher tendencies separately for each season. However, xCTRL is flexible and can be applied over any justifiable time frame. The visuals presented above reflect tendencies from the 2023 MLB season; we make no claims about their validity outside that range.

\begin{figure}[htb!]
    \centering
    \subfloat[]{{\includegraphics[height=0.35\textwidth]{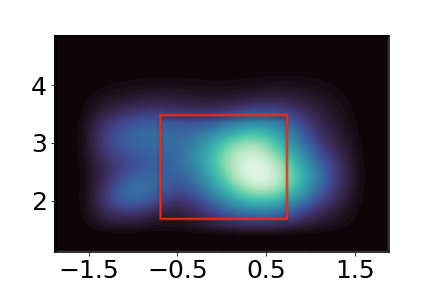}}\label{fig:verlander_heatmap_2011}}%
    \subfloat[]{{\includegraphics[height=0.35\textwidth]{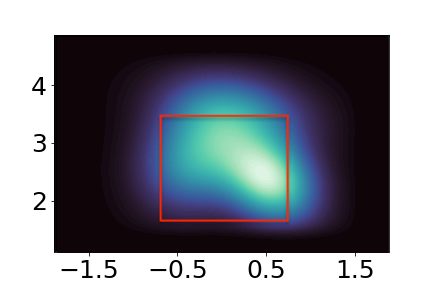}}\label{fig:verlander_heatmap_2019}}%
\caption{
  Justin Verlander's pitch location density for four-seam fastballs against right-handed batters in 2011 (a) and 2019 (b).
}
    \label{fig:verlander_heatmap}
\end{figure}

Justin Verlander's career illustrates this point clearly. He has been in the league for nearly two decades and has consistently relied on his fastball. To see how his strategy has evolved, we visualize in Figure~\ref{fig:verlander_heatmap} his fastball location tendencies in 2011 and 2019, the seasons in which he won his first two Cy Young Awards. 
In 2011, Verlander's targets were primarily lateral. By 2019, he favored high fastballs with less emphasis on inside locations. Capturing these shifts is one of xCTRL’s strengths: it infers intent individually for each season, so a deliberate change in policy is not treated as a failure of control.

\begin{figure}[htb!]
    \centering
    \subfloat[]{{\includegraphics[height=0.35\textwidth]{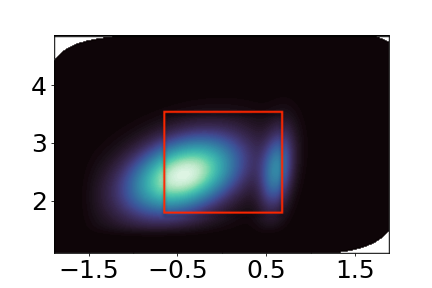}}\label{fig:heatmap_Steele10FF}}%
    \subfloat[]{{\includegraphics[height=0.35\textwidth]{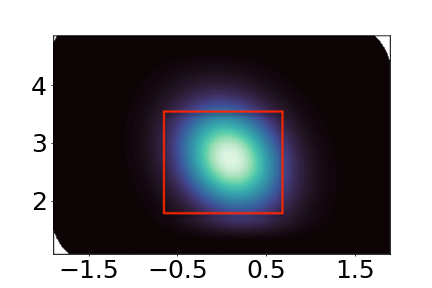}}\label{fig:heatmap_Ryan10FF}}%
\caption{
   Pitch location densities for Justin Steele (a) and Joe Ryan's (b) four-seam fastball against right-handed batters in 2023 in a 1-0 count (n = 151 \& 76, respectively).
}
    \label{fig:heatmaps_10count}
\end{figure}

Location policy may also depend on game-state. A natural starting point is to examine how pitch targeting changes by count. While there is often limited data to identify precise tendencies for every count with high confidence, we illustrate in Figure~\ref{fig:heatmaps_10count} the general principle using fastball distributions in 1-0 counts.
Even in the same count, pitchers may adopt very different strategies. For example, Justin Steele targets two distinct zones––low-and-inside and middle-away––while Joe Ryan tends to throw directly over the middle. Both pitchers score well under xCTRL. However, traditional control metrics like Location+ assume a uniform league-wide response to count and would penalize one of these pitchers for deviating from the expected trend. This highlights a key advantage of xCTRL: it treats consistent, pitcher-specific behavior as intentional rather than erroneous. Conscious deviations from the norm are accepted and appropriately factored into control measurement.

Unfortunately, there is not enough data to reliably analyze count-specific intent at scale. In Section~\ref{sec:moreAnalyses} we introduce a smoothing approach using count types and in Appendix~\ref{app:weighted_em_count} we propose a general extension to xCTRL for low-data settings.

\begin{figure}[hbt!]
\centering
\includegraphics[width=0.5\textwidth]{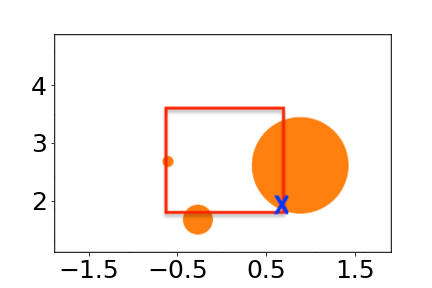}
\caption{
  Estimated location intentions for Max Fried low-and-away fastball against right-handed batter
}
\label{fig:fried_intentions}
\end{figure}

After a pitch is thrown, xCTRL updates intent estimates via a posterior adjustment. Figure~\ref{fig:fried_intentions} provides a visual example of this update for a single pitch thrown by Max Fried. We begin with Fried's fastball location tendencies against right-handed batters (Figure~\ref{fig:friedrhb}). Consider a pitch that lands low-and-away, marked by a blue X. Using Bayes rule, xCTRL assigns posterior probabilities to each of Fried’s three typical targets: $(0.90,\ 0.01,\ 0.09)$. These values are visualized by circle sizes in the figure.
The Euclidean distances from the pitch to each target center are $(8.40,\ 17.82,\ 11.33)$ inches. xCTRL for this pitch is the dot product of the posterior probability vector and the distance vector, yielding a control score of $8.76$ inches. This is a decent xCTRL score. The pitch could plausibly be a vertical miss from the middle-away target or a horizontal miss from the low-and-inside target. xCTRL identifies middle-away as the most likely intended location because it has a higher prior weight and a shorter execution distance.
Although the pitch does not land exactly on any target, it aligns well with Fried’s established fastball patterns to right-handed hitters. In other words, it is considered ``on target'' not because it conforms to a league-wide expectation, but because it was consistent with Fried's individualized intent.

Given sufficient data, we can generate these visuals for any pitcher and any pitch. They provide insight into a pitcher's location policy and how it may vary across situations. After a pitch is thrown, the posterior update allows us to visualize the updated intent distribution, offering an intuitive view of how xCTRL is calculated for that pitch. Together, these visuals illustrate the core process behind calculating xCTRL: establishing individualized location tendencies and refining intent estimates based on the actual pitch location.

\subsection{Comparing pitcher rankings by xCTRL and Location+}\label{sec:control_ranking}

We compute the average xCTRL score across all pitches in a given bin $\bx$, where we bin by pitcher, season, pitch type, and batter handedness.
In this section, we report a pitcher's overall xCTRL as the average between his control against left- and right-handed batters. 
While pitch tendency distributions differ by handedness, the resulting xCTRL magnitudes are generally similar. The rankings given here are based on fastball control, as control is often considered more important for fastballs than for other pitch types. Rankings for other pitch types can be computed in the same way.

We use these rankings as a preliminary validation of xCTRL. In general, we expect xCTRL to align with conventional baseball intuition and established control metrics. To assess this, we compare xCTRL to Location+, which serves as our baseline for existing location-based control estimates. We collect all starting pitcher-seasons from 2021–2023 for which both xCTRL and Location+ are available, yielding 44 pitcher-seasons in 2023 and 118 overall.

\begin{figure}[htb!]
    \centering
    \subfloat[]{{\includegraphics[width=0.15\textheight]{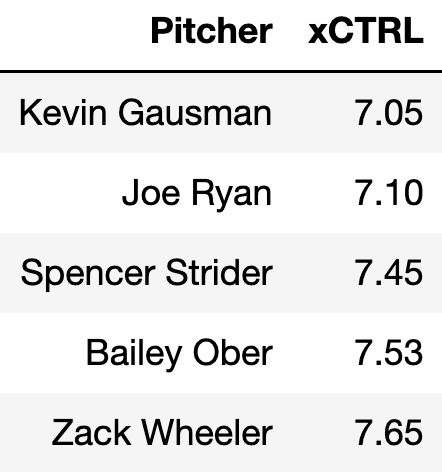}}\label{fig:bestControl2023}}
    \qquad
    \subfloat[]{{\includegraphics[width=0.13\textheight]{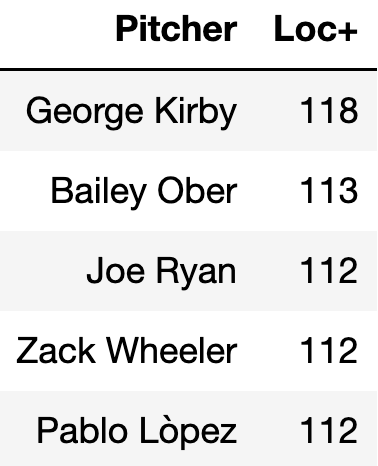}}\label{fig:bestLoc2023}}
    \qquad
    \subfloat[]{{\includegraphics[width=0.15\textheight]
    {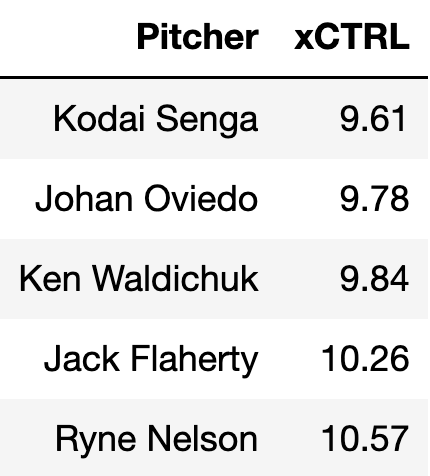}}\label{fig:worstControl2023}}
    \qquad
    \subfloat[]{{\includegraphics[width=0.15\textheight]
    {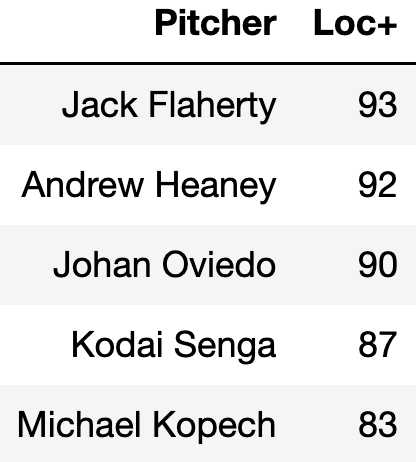}}\label{fig:worstLoc2023}}
\caption{
  Top five fastball pitchers in 2023 by control (a) and Location+ (b). Bottom five fastball pitchers in 2023 by control (c) and Location+ (d)
}
    \label{fig:pit_rankings_2023}
\end{figure}

\begin{figure}[htb!]
    \centering
    \subfloat[]{{\includegraphics[width=0.17\textheight]{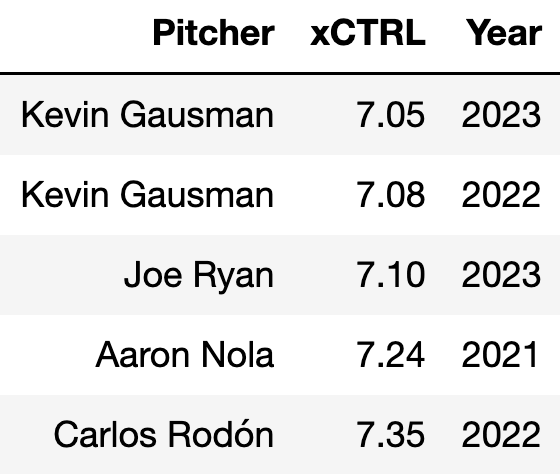}}\label{fig:bestControlEver}}
    \
    \subfloat[]{{\includegraphics[width=0.17\textheight]{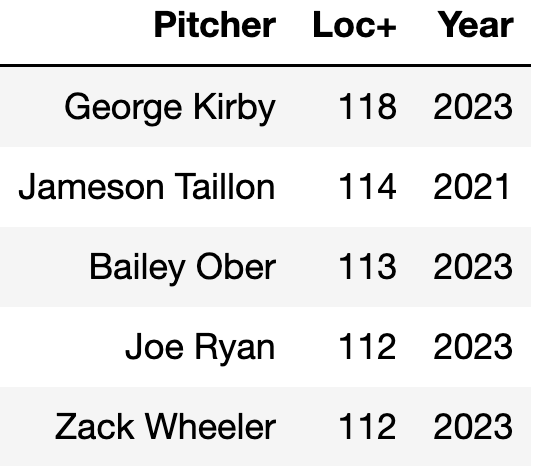}}\label{fig:bestLocEver}}
    \
    \subfloat[]{{\includegraphics[width=0.18\textheight]
    {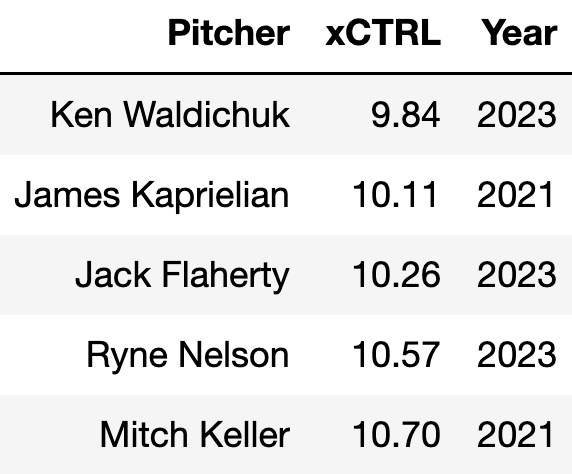}}\label{fig:worstControlEver}}
    \
    \subfloat[]{{\includegraphics[width=0.17\textheight]
    {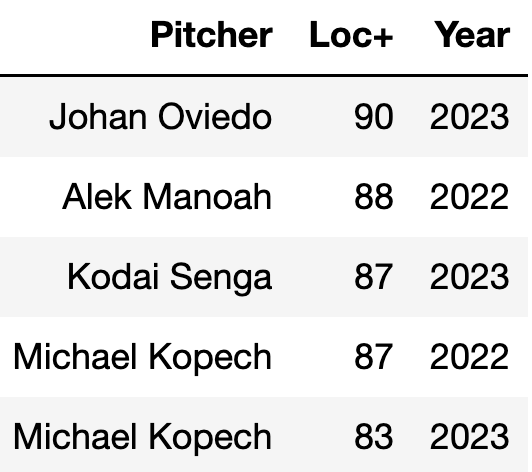}}\label{fig:worstLocEver}}
\caption{
  Top five fastball pitchers from 2020-2023 by control (a) and Location+ (b). Bottom five fastball pitchers from 2020-2023 by control (c) and Location+ (d)
}
    \label{fig:pit_rankings_total}
\end{figure}

In Figure~\ref{fig:pit_rankings_2023} we display the top and bottom five pitchers in 2023 by both xCTRL and Location+. In Figure~\ref{fig:pit_rankings_total} we display the top and bottom five pitcher-seasons from 2021–2023. xCTRL is measured in inches, with lower values indicating better control. Location+ is a standardized metric where 100 is league average and higher values reflect better control.

\begin{figure}[hbt!]
\centering
\includegraphics[width=0.5\textwidth]{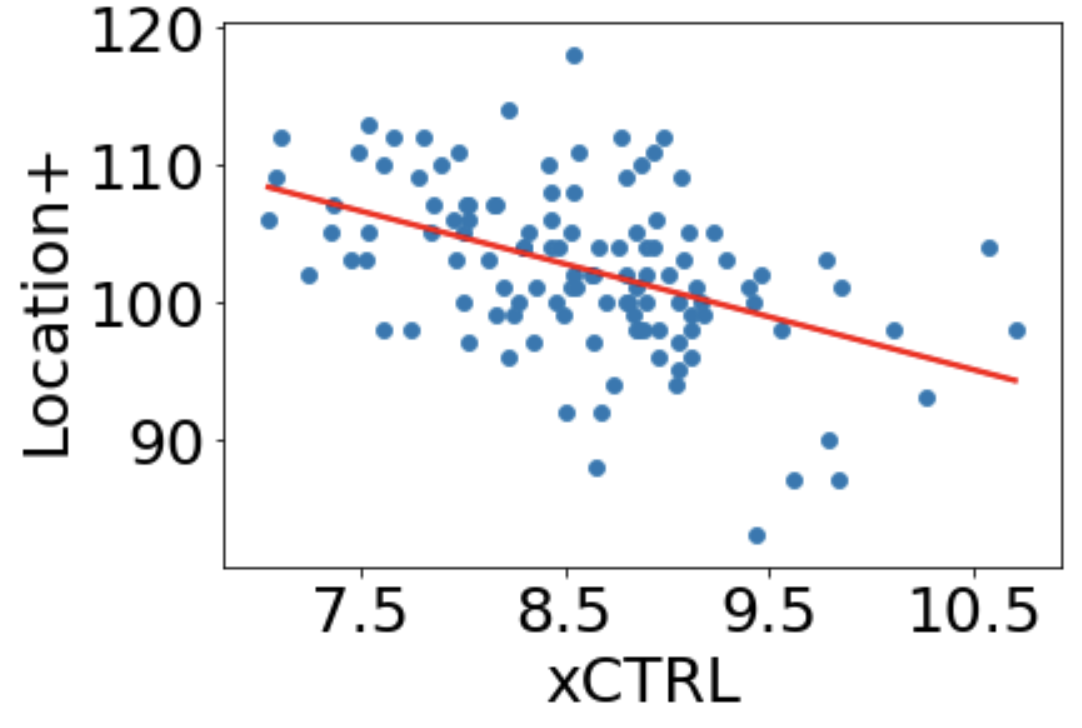}
\caption{
  Location+ ($y$-axis) versus xCTRL ($x$-axis) for pitcher-season fastballs against right-handers.
}
\label{fig:controlLocation}
\end{figure}

xCTRL and Location+ generally produce similar rankings for fastball control. Pitchers who rate highly under xCTRL typically grade well under Location+ too. This relationship is reflected in the moderately strong correlation ($r = -0.46$) shown in Figure~\ref{fig:controlLocation}. However, there are notable exceptions. For example, xCTRL consistently rates Kevin Gausman's fastball control as elite, ranking his 2021 and 2023 seasons as the top two pitcher-seasons in fastball control since 2021. In contrast, Location+ views his control as solid but not extraordinary, placing him $13^{th}$ out of $44$ starting pitchers in 2023. We explore the importance of this discrepancy in the next section.

The units of xCTRL offer an intuitive sense of what we mean by ``control.'' Kevin Gausman's xCTRL score is remarkable: he missed his intended target by an average of just 7.05 inches. A baseball is slightly less than 3 inches in diameter, meaning the best control pitchers miss by just over two baseball widths on average. Even the worst control pitchers in our rankings miss by only slightly more than three baseball widths. Considering the velocity and difficulty of each pitch, these levels of precision are impressive. 

These xCTRL rankings also serve as a sanity check. Pitchers with strong reputations for control generally appear at the top (e.g. Gerrit Cole), while those known for being erratic are near the bottom (e.g. Dylan Cease) \footnote{
    Pitchers categorized as swing-and-miss vs location-driven  from the 2023 season \url{https://blogs.fangraphs.com/jacob-degrom-isnt-like-other-pitchers/}
}. Moreover, year-over-year consistency in xCTRL rankings suggests that it is a stable and reliable measure of pitcher control (see the end of Section~\ref{sec:model_validation} for a more detailed analysis of the stability of xCTRL across seasons.). 

\subsection{Validating our control metric}\label{sec:model_validation}

xCTRL rankings broadly align with pitcher success, suggesting the metric captures meaningful performance differences. Three of the top five pitcher-seasons by xCTRL from 2021-2023 received Cy Young votes, while none of the bottom five had an ERA below 4.00. To formally evaluate the predictive value of xCTRL, we compare the strength of xCTRL and Location+ for predicting season-level outcomes, controlling for Stuff+ (a measure of pitch quality based on physical characteristics such as velocity, movement, and release point \citep{stuffprimer}). We focus on fastball xCTRL and restrict our dataset to starting pitcher-seasons from 2021--2023 for which xCTRL, Location+, and Stuff+ are all available. While season-level analysis includes some confounding factors—since performance reflects full repertoire and game-level decisions—it allows for a fair comparison between xCTRL and Location+ under consistent conditions. We present a more granular evaluation in the next section.

\begin{figure}[hbt!]
\centering
\includegraphics[width=0.75\textwidth]{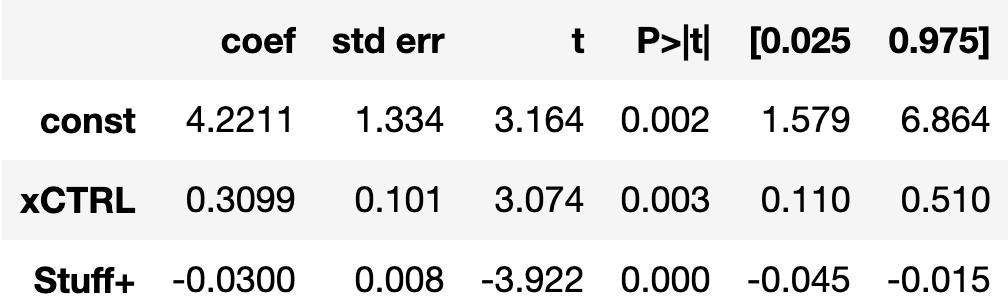}
\caption{
  Regression summary for regressing FIP on xCTRL and Stuff+.
}
\label{fig:FF_control_FIP}
\end{figure}

We begin by predicting the highest-level outcome: run prevention. Because ERA is known to be noisy, we use Fielding Independent Pitching (FIP) as our dependent variable. Specifically, we regress FIP on xCTRL and Stuff+ (see Figure~\ref{fig:FF_control_FIP}). Both xCTRL and Stuff+ are statistically significant. The sign of each coefficient is as expected. A higher xCTRL score (indicating worse control) is associated with higher FIP (a worse pitching outcome), while a higher Stuff+ score (indicating better pitch quality) predicts lower FIP. We estimate that improving fastball xCTRL by one inch, holding Stuff+ constant, leads to an expected reduction of about $0.3$ FIP. Since FIP is on the same scale as ERA, this effect size is substantial. Given that xCTRL scores span a roughly three-inch range between the best and worst fastball control pitchers, this suggests that control alone could explain about a full run difference in FIP between those pitchers, assuming similar pitch quality. That’s roughly one run saved per game due to fastball control alone.

We also estimate that a one-SD increase in Stuff+ (which has mean $\approx 100$ and SD $\approx 10$) leads to an expected reduction of about $0.3$ FIP––comparable in magnitude to the effect of a one-inch improvement in xCTRL (mean $\approx 9.14$ and SD $\approx 1.10$). Though they capture distinct pitcher attributes, both pitch quality and control have similarly meaningful impacts on run prevention. A one-SD increase in Stuff+ and xCRTL both represents a similar improvement in reducing runs allowed as measured by FIP.

\begin{figure}[hbt!]
\centering
\includegraphics[width=0.75\textwidth]{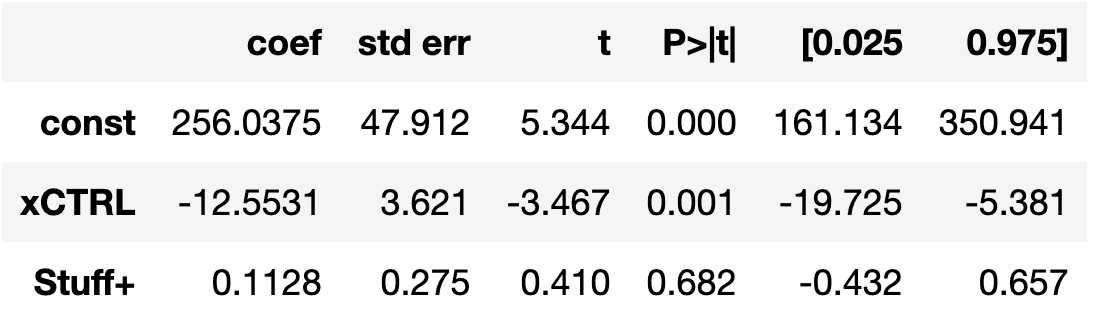}
\caption{
  Regression summary for regressing Innings Pitched on xCTRL and Stuff+.
}
\label{fig:FF_control_IP}
\end{figure}

Pitchers with excellent pitch quality but poor control are often seen as more volatile and less reliable as starters. If that’s true, does xCTRL help predict pitcher reliability over a season? To explore this, we regress Innings Pitched (IP) on fastball xCTRL and Stuff+ (see Figure~\ref{fig:FF_control_IP}). xCTRL is strongly statistically significant, while Stuff+ is not significant. In a similar regression with Location+ in place of xCTRL, Location+ is also not significant, providing evidence that xCTRL is a stronger indicator of season-long reliability than Location+. Hence, xCTRL is not just useful for predicting rate-based metrics like FIP, but also for predicting cumulative pitcher value. 

\begin{figure}[htb!]
    \centering
    \subfloat[]{{\includegraphics[height=0.23\textwidth]{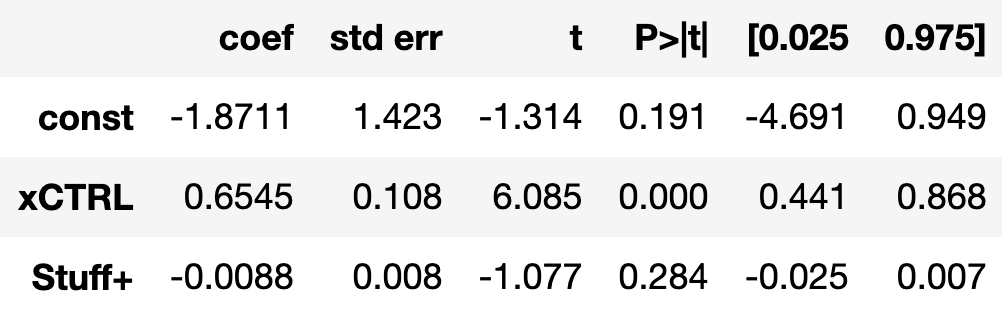}}}\label{fig:bb9_reg}%
    \qquad
    \subfloat[]{{\includegraphics[height=0.23\textwidth]{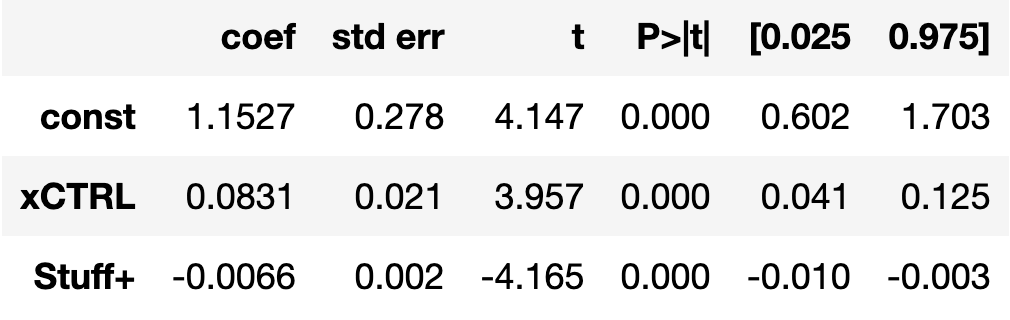}}}\label{fig:whip_reg}%
\caption{
  Regression summary for regressing BB/9 (a) and WHIP (b) on xCTRL and Stuff+.
}
    \label{fig:bb9whip}
\end{figure}

On average, pitchers with elite fastball control are expected to throw about 36 more innings per season than those with poor control (holding Stuff+ constant). This effect may be understated––our dataset includes only starting pitchers with enough fastballs to compute xCTRL, excluding many who missed time due to injury. Stuff+ is positively correlated with injury risk \citep{stuffinjuries}, and we find that xCTRL is negatively correlated with Stuff+ ($r = -0.34$). Thus, high-control, low-Stuff+ pitchers may accumulate even more innings due to greater durability––an effect not fully captured here.

xCTRL is also a significant predictor of more traditional outcome-based measures of control In Figure~\ref{fig:bb9whip}, we show that xCTRL significantly predicts both BB/9 and WHIP. Pitchers with poor xCTRL are predicted to give up more walks, while Stuff+ is not a significant predictor. We predict the worst control pitchers to allow roughly two additional walks per game compared to the best. 
Interestingly, both xCTRL and Stuff+ are significant predictors of WHIP, suggesting that xCTRL captures both walk and hit prevention, whereas Stuff+ is primarily associated with limiting hits. 
Similar regression with Location+ in place of xCTRL reveal that Location+

Location+ shows similar significance levels to xCTRL for both BB/9 and WHIP, but fails to significantly predict control-related outcomes across all control-related measures, such as Innings Pitched. In contrast, xCTRL is a significant predictor in every control-related regression we test, reinforcing its value as a reliable and general-purpose control metric.

\begin{figure}[htb!]
    \centering
    \subfloat[]{{\includegraphics[width=0.21\textheight]{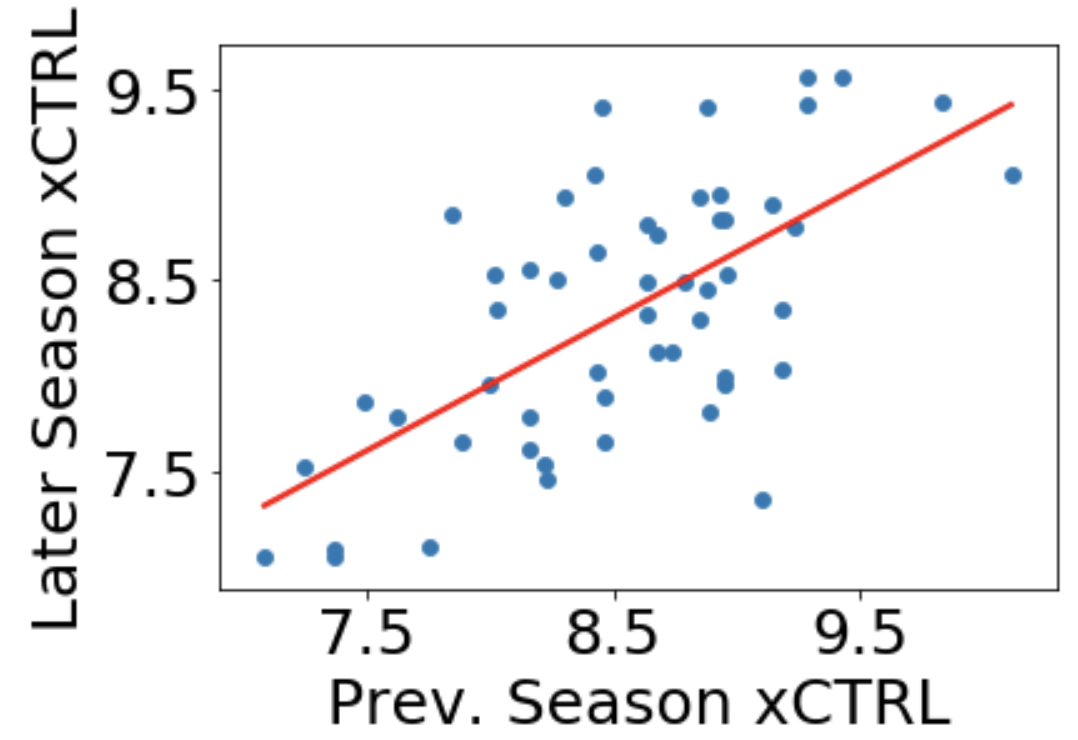}}\label{fig:xCTRLYears}}
    \qquad
    \subfloat[]{{\includegraphics[width=0.21\textheight]{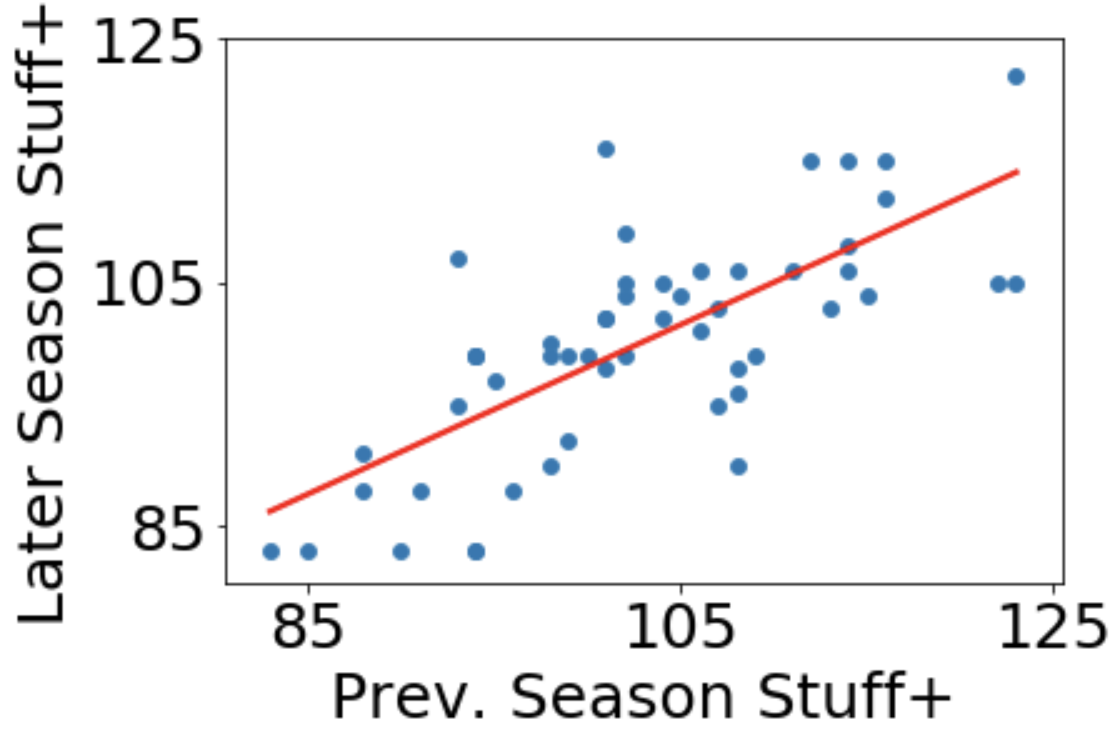}}\label{fig:stuffYears}}
    \qquad
    \subfloat[]{{\includegraphics[width=0.21\textheight]
    {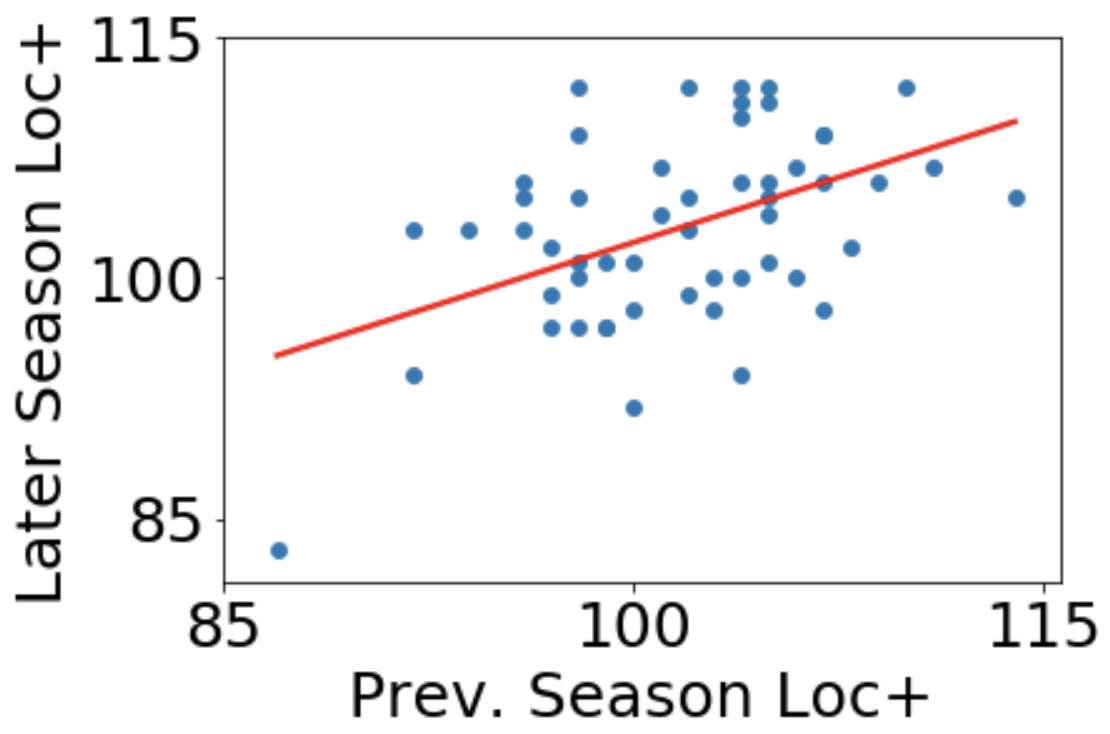}}\label{fig:locYears}}
\caption{
  Scatterplots comparing results from a previous to following season for fastball xCTRL (a) Stuff+ (b) and Location+ (c). Correlations are 0.65 (a), 0.73 (b), 0.48 (c)
}
    \label{fig:year_over_year}
\end{figure}

Finally, we assess the stability of xCTRL across seasons. Among pitchers with multiple qualifying seasons, we observe an inter-season correlation of 0.65 for fastball xCTRL ~\ref{fig:year_over_year}, based on 53 pitcher-season pairs. For comparison, Stuff+ shows slightly higher year-over-year stability at 0.73, while Location+ is considerably less stable at 0.48. Notably, ERA has an inter-season correlation of just 0.19. These results indicate that xCTRL is not only predictive of performance but also reflects a stable pitcher trait, further supporting its validity as a measure of control.

\subsection{ {More granular analyses enabled by pitch-level control} }\label{sec:moreAnalyses}

xCTRL advances control measurement not only by offering stronger predictive power, but also by enabling more granular and flexible analysis. Metrics that rely on hidden or proprietary computations—such as Location+ and Command+—can only be analyzed at the level of detail made publicly available. For Location+, this typically means season-level or, at best, game-level summaries. In contrast, xCTRL is computed at the individual pitch level. It can be aggregated to any level of interest by averaging control scores within a chosen bin $\bx$, allowing for situational analyses that were previously infeasible.

To illustrate this analytical power, we investigate the impact of a single poorly located pitch––an analysis not possible with Location+. Because the run value of a pitch outcome depends on game context (e.g., a single with runners in scoring position is more costly than one with the bases empty), we use RE24 to measure the value of each at-bat outcome \citep{RE24}. For all at-bats thrown thrown by a pitcher in our dataset from 2021–2023 that end in a fastball ($n = 23,831$), we compute the xCTRL of that final pitch (fastball). Since Stuff+ is not available at the pitch level, we use velocity as a proxy for pitch quality. 

\begin{figure}[hbt!]
\centering
\includegraphics[width=0.75\textwidth]{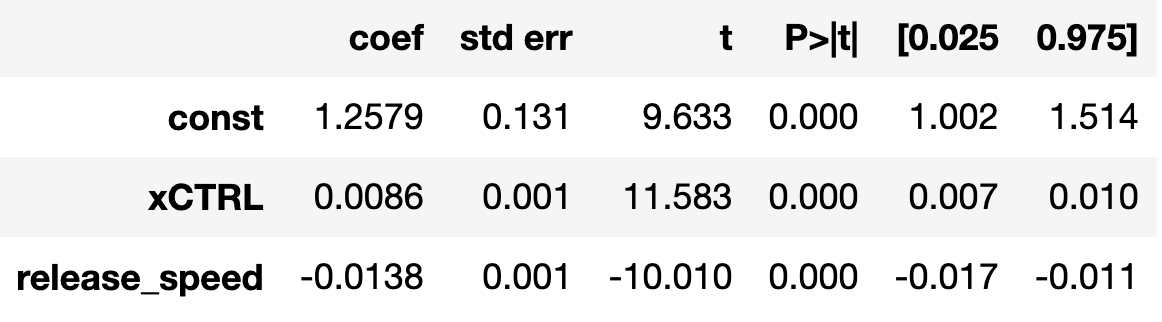}
\caption{
  Regression summary for regressing RE24 on xCTRL and fastball velocity.
}
\label{fig:re24_reg}
\end{figure}

We then regress RE24 on xCTRL and velocity (see Figure~\ref{fig:re24_reg}). Both fastball xCTRL and velocity are highly significant predictors of runs allowed on a per-pitch basis. Although the coefficients are small due to the fine-grained level of analysis, their implications are meaningful: control alone accounts for a difference of roughly half a run per start between the best and worst control pitchers, assuming $20$ batters faced per outing (since $0.0086\cdot3\cdot20 \approx 0.5)$. Every additional mile-per-hour of fastball velocity is expected to reduce runs by $\approx 0.25$ runs over the same $20$ batter period. 

xCTRL also enables analysis at the individual pitch level, regardless of whether a pitch was put in play. We use the change in expected runs before and after a pitch ($\Delta RE$, which is included in Statcast data) to reflect the success of a pitch. Unlike RE24, $\Delta RE$ captures the value of balls, strikes, and other intermediate outcomes, allowing us to include every pitch in the analysis, not just those ending an at-bat. 

We use this setting to assess the value of control on non-fastballs, focusing on curveballs. Off-speed pitch usage can be highly situational; for instance, a pitcher might throw a curveball early in the count to steal a strike, rather than to generate a whiff. Measuring the value of control across all pitch contexts helps capture this variation. We calculate xCTRL for every curveball thrown by a qualified pitcher during the 2023 season, resulting in $n = 6,280$ pitches. Because Stuff+ is not available at the individual pitch level, we use spin rate as a proxy for pitch quality.

\begin{figure}[hbt!]
\centering
\includegraphics[width=0.75\textwidth]{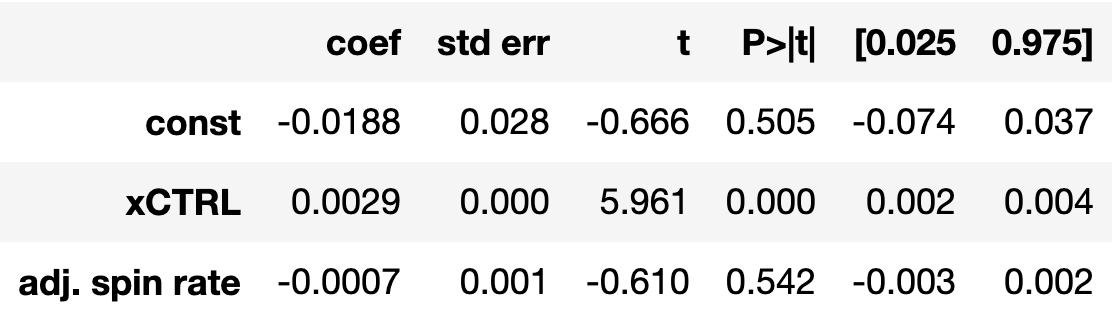}
\caption{
  Regression summary for regressing $\Delta RE$ on curveball xCTRL and spin rate (adjusted to represent rotations per $1/2$ second).
}
\label{fig:dre_CU}
\end{figure}

We then regress $\Delta RE$ on xCTRL and spin rate (see Figure~\ref{fig:dre_CU}). Curveball xCTRL is a significant predictor of pitch success. Specifically, an improvement of one inch in curveball xCTRL leads to 0.29 fewer expected runs allowed per 100 curveballs, holding spin rate constant. This result is consistent with baseball intuition: a poorly located curveball that doesn't break as intended––commonly referred to as a ``hanger''––is one of the easiest pitches to hit. Repeatedly missing with curveballs increases the likelihood of hangers, and with it, the risk of giving up runs.

xCTRL's flexibility enables more detailed analysis by allowing users to define bins tailored to any situation of interest. One natural use case is examining how a pitcher’s location tendencies evolve with the count. To explore this, we group counts into three categories––early, hitter-friendly, and pitcher-friendly––to ensure each group contains enough data to produce stable tendency distributions.

\begin{figure}[htb!]
    \centering
    \subfloat[]{{\includegraphics[height=0.3\textwidth]{WheelerRHBFF2023.png}}}\label{fig:overallWheel}%
    \qquad
    \subfloat[]{{\includegraphics[height=0.3\textwidth]{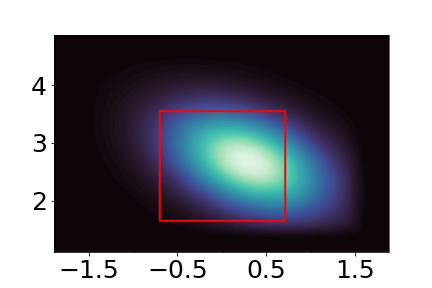}}}\label{fig:earlyWheel}%
    \qquad
    \subfloat[]{{\includegraphics[height=0.3\textwidth]
    {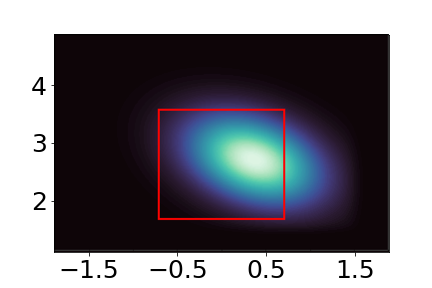}}}\label{fig:hitterWheel}%
    \qquad
    \subfloat[]{{\includegraphics[height=0.3\textwidth]
    {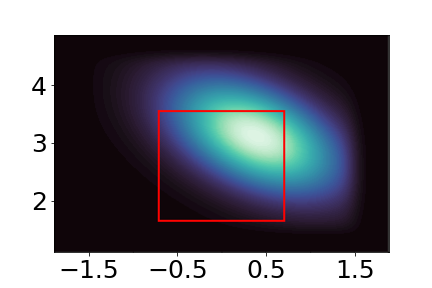}}}\label{fig:pitcherWheel}%
\caption{
  Zack Wheeler's four-seam fastball pitch location density against right-handed batters across all counts (a), in early counts (0-0, 0-1, 1-0, 1-1) (b), hitter-friendly counts (2-0, 2-1, 3-1) (c), and pitcher-friendly counts (0-2, 1-2, 2-2) (d).
}
    \label{fig:count_type_wheel}
\end{figure}

Zack Wheeler’s fastball location policy against right-handed batters offers a clear example of count-dependent targeting (see Figure~\ref{fig:count_type_wheel}). Overall, Wheeler has two primary targets: middle-middle and middle-away. In both early and hitter-friendly counts, he tends to stay safely within the strike zone, with only slight tightening of his target range in hitter-friendly situations. However, in pitcher-friendly counts, Wheeler shifts his strategy and frequently targets up-and-away––a location consistent with conventional baseball wisdom. When ahead in the count, pitchers are more willing to aim at the edges of the zone, where a potential ball is less costly. 

Crucially, xCTRL does not assume all pitchers follow this pattern. Some may change their targets based on count, while others may not adjust at all. By estimating individualized location tendencies, xCTRL avoids imposing uniform behavioral assumptions and instead evaluates each pitcher on his own terms.

Count is just one of many contextual factors that can influence location policy. A pitcher may adjust based on the batter’s power profile or whether there are runners in scoring position. Thanks to xCTRL's flexible binning, any of these scenarios can be analyzed, provided sufficient data. For cases with limited data, we propose a smoothing-based extension to the xCTRL algorithm in Appendix~\ref{app:weighted_em_count}.

\section{Discussion}\label{sec:discussion}

Control has long been understood within the baseball community as an important indicator of a pitcher's success, yet quantifying it has proven challenging. Existing control metrics often rely on simplifying and unfounded assumptions; pitcher individuality should be embraced, not obscured. With the advent of precise pitch location data, we can form more accurate beliefs about pitcher intent. In this paper, we introduce xCTRL as a principled and interpretable solution for quantifying control. Our calculations and code are publicly available for use by the broader baseball community (see Appendix~\ref{app:code}).

xCTRL can be applied to a wide range of pitcher analyses beyond those presented here. Fundamentally, xCTRL is the average pitch-level control within a specified bin $\bx$. In this paper, we report xCTRL by pitch type for each pitcher-season. Bins can be customized to be as specific or general as needed. For instance, one could analyze how Gerrit Cole locates his fastball against power hitters versus contact hitters. A front office could track the development of a pitching prospect by monitoring year-over-year changes in xCTRL. Even batters could use xCTRL visualizations to study a pitcher's most common targets and their frequencies. The flexibility in bin selection makes xCTRL a powerful tool for diverse analytical purposes.

That said, xCTRL is not without limitations. Bins that are too specific may suffer from insufficient data, making it difficult to fit stable tendency distributions. Additionally, care must be taken to ensure that the data used reflects a single, coherent location policy. Fitting a model to data that blends multiple strategies may lead to misleading results. We address some of these concerns with an extension for low-data settings in Appendix~\ref{app:weighted_em_count}.

Another compelling aspect of xCTRL is its transparency. It offers an intuitive, interpretable measure: the distance (in inches) between a pitch's actual location and its inferred target. If we want to see where the pitcher was likely aiming, we can visualize it. Every step of the computation is verifiable––there is no black box requiring blind trust. The results are conclusive: xCTRL is a stable, informative control metric with significant predictive value for pitcher success and usage. By adopting a more mathematically rigorous yet conceptually intuitive approach––measuring location relative to a pitcher’s own tendencies––we substantially improve upon existing control metrics.

\if0\blind
{
  \section*{Acknowledgments}
  The authors thank Abdelrahman Mohamed for his help with navigating preexisting software packages.
} \fi

\bibliography{refs}

\clearpage
\newpage
\begin{center}
{\large\bf SUPPLEMENTARY MATERIAL}
\end{center}
\appendix

\section{Accessing the Code and Data}\label{app:code}

Our code is available on Github at {\href{https://github.com/mattludwig6/Pitching-Control-Metric}{https://github.com/mattludwig6/Pitching-Control-Metric}.
We collected Statcast pitch data using the pybaseball package \url{https://github.com/jldbc/pybaseball}. 
The ``Pitch Rankings'' folder gives the median control score and associated intent distribution for every qualified pitcher, pitch type, and handedness of batter combination. 
The other files represent the code needed to compute this control metric for any future scenario, our initial attempt at the approach in Appendix~\ref{app:weighted_em_count}, and the initial analysis of control that led us down this path.

\section{Control in Specific Cases}\label{app:weighted_em_count}

We'd like to estimate control given a larger covariate set $\bx$, i.e. for combinations of pitcher, season, pitch type, batter handedness, and count.
It is possible that pitchers employ differing strategies depending on the count, which could lead to different control values.
The challenge is these bins can leave us with too small a sample size to simply run the EM algorithm from Section~\ref{sec:em}. 
Hence, in this section we propose a modification of that algorithm to shrink a count-specific pitch location density towards the count-agnostic one. 

To do so, we use Bayesian nonparametric learning, in particular Algorithm 3 from \citet{npl}.
The count-agnostic pitch location density, fit using the EM algorithm from Section~\ref{sec:em}, acts as the prior.
We generate 250 synthetic pitch locations from the prior.
Then we combine the observed count-specific pitch locations with the synthetic ones, downweighting the synthetic data. The synthetic pitch location sample weights, $\omega \in [0,1]$, determine how much the count-specific pitch location density shrinks to the count-agnostic one. As $\omega$ gets larger, the count-specific density approaches the count-agnostic prior. To determine $\omega$, we create a mesh of the unit interval and test several sample weights $[\omega_{1},\ldots,\omega_{n}]$. For a particular train/test split, we run the EM with fixed weights from \citep{weightedem} and record the resulting log-likelihood. We choose the $\omega$ that minimizes the negative log-likelihood of the data. After determining $\omega$, we run our algorithm to produce $B=100$ pitch location densities. We create a confidence interval of control values and select the median pitch location density from these results.

Since the resulting density depends on random seeds and randomly generated pseudo-samples, the EM with Fixed Weights may get stuck at different local minima for a Gaussian Mixture. To account for this, we use $R=5$ random restarts each run through our algorithm. From each set of restarts, we select the density with the highest out-of-sample likelihood and add that to our set of $B$ pitch location densities.


We applied this algorithm to estimate Cole's intent for fastballs against right-handed batters in early counts 0-0, 0-1, and 1-0.
In Figure~\ref{fig:cole_heatmap_weighted} we visualize the fitted tendency distribution for two different weightings of pseudo-samples from the count-agnostic intent distribution.
As the synthetic data weight approaches one, the count-specific intent distribution approaches the count-agnostic one. When $\omega$ is small, it's easy to see the target identified from count-specific data alone. A new, low frequency target is visualized near the middle of the strike-zone. Meanwhile the predominant target is diagonal near the top of the zone, representing Gerrit Cole's overall tendency to throw along the top and outside edges of the strike-zone. As $\omega$ gets larger, count-specific tendencies look more similar to his overall targets. There should be an optimal weight that shrinks the right amount.

\begin{figure}[htb!]
    \centering
    \subfloat[]{{\includegraphics[height=0.35\textwidth]{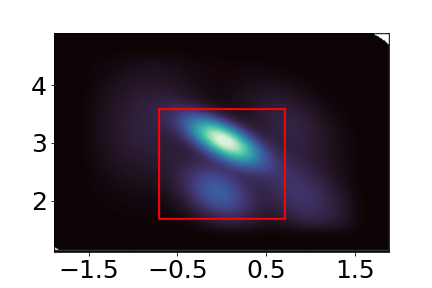}}\label{fig:cole_heatmap_w0.1}}%
    \subfloat[]{{\includegraphics[height=0.35\textwidth]{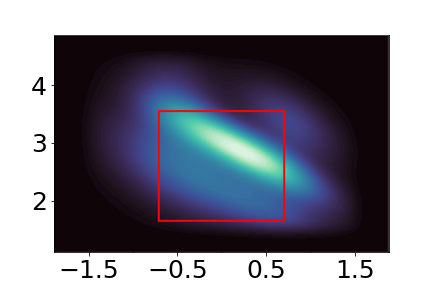}}\label{fig:cole_heatmap_w0.7}}%
\caption{
  Heatmap of 2023 Gerrit Cole's pitch location density for fastballs against right-handed batters in counts 0-0, 0-1, and 1-0 with synthetic data weight $\omega = 0.1$ (a) and $\omega = 0.7$ (b).
}
    \label{fig:cole_heatmap_weighted}
\end{figure}

Ultimately, we feel that the count-specific intent distributions aren't too trusthworthy.
We saw some oddities in the results that make us apprehensive to accept the results in their current state.
For example, there are many discrepancies between the overall fastball control rankings and the rankings in a 1-2 count, such as Joe Ryan who went from top five to average and Tanner Bibbee who went from bottom five to average.
The lack of data in each bin presents a challenge that we leave to fully solve in future work.

\section{Simulation of Control}\label{sec:sims}

To illustrate the value of pitcher control, we conduct a simulation study. We use real game fastball data from Jacob deGrom and Mike Trout's careers to represent a pitcher-batter interaction. Our data is in a 5x5 discretized extended strike-zone, which we got from \citep{zonedata} and visualized in Figure~\ref{fig:JD}. We average the values by zone between Jacob deGrom and Mike Trout's heat maps to make a single set of heat maps representing this pitcher-batter interaction. From this set of heat maps, we create a probability vector representing the likelihood of all outcomes when Jacob deGrom throws a fastball to Mike Trout. We compute 25 such probability vectors, one for each box of the strike-zone. We sample from these probability vectors and follow standard baseball rules to simulate an inning.

To investigate the value of pitcher control, we choose an intended location and add white noise in the horizontal and vertical directions to determine where the pitch will cross the strike-zone plane. The white noise is represented by i.i.d samples from $\mathcal{N}(0,\sigma_{T})$, where we set $\sigma_{T}$. $\sigma_{T}$ represents control, where smaller $\sigma_{T}$ leads to less deviation between intended and actual pitch locations. For each value of $\sigma_{T}$, we simulate 10,000 innings and record the runs scored. We average the runs scored over 10,000 simulated innings and show how it varies according to $\sigma_{T}$ in Figure~\ref{fig:expectedRuns}.

\begin{figure}[hbt!]
\centering
\includegraphics[width=0.65\textwidth]{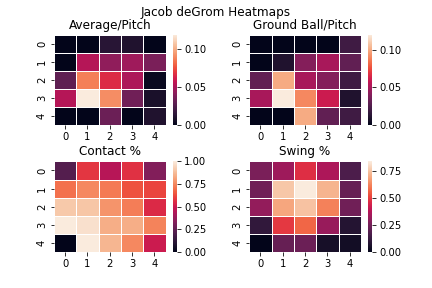}
\caption{
  Heatmaps representing outcome values for Jacob deGrom in each part of the discretized extended strike-zone
}
\label{fig:JD}
\end{figure}

We notice that control and expected earned runs have a non-linear relationship. This implies that improving control provides different overall value depending on the value of a pitcher's original control. This reaffirms the importance of measuring a pitcher's control independent of the rest of the league. 

\begin{figure}[hbt!]
\centering
\includegraphics[width=0.65\textwidth]{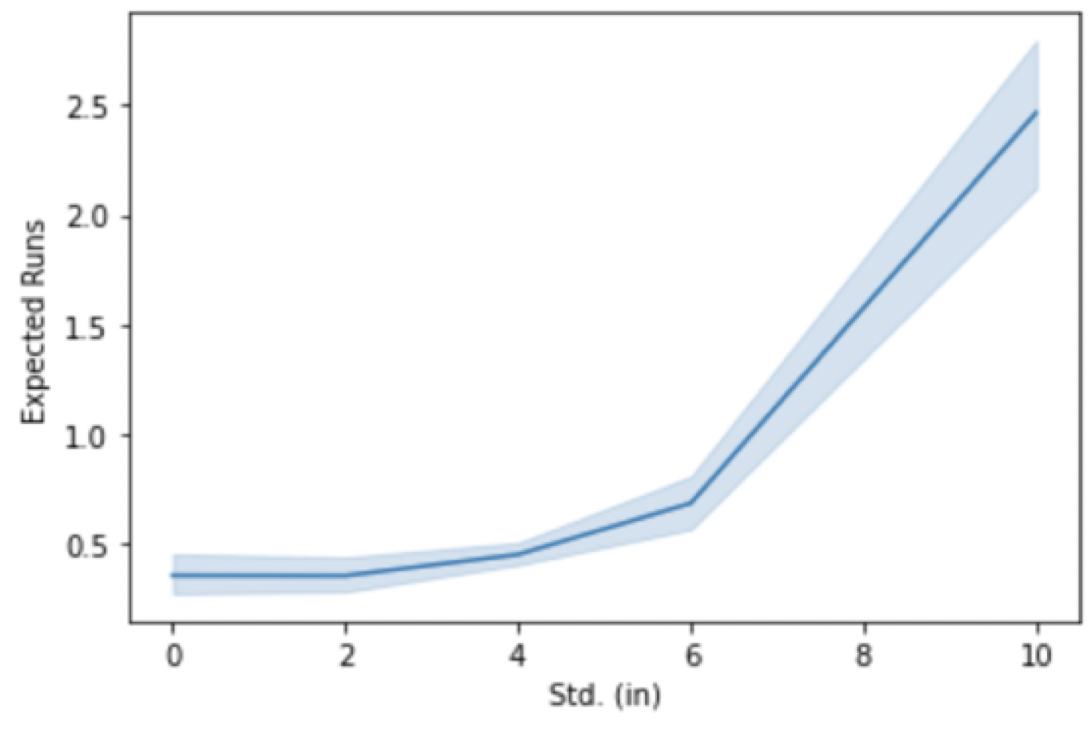}
\caption{
  Average runs per inning in simulation as a function of pitcher control
}
\label{fig:expectedRuns}
\end{figure}

Following this, we explore how a pitcher's perception of their control may impact earned runs. To do this, we define an optimal location policy which is a function of game state and a pitcher's control. In general in this simulation, a pitcher with weak control minimizes their expected runs by intending to throw closer to the middle of the strike-zone.

We start with a pitcher that has $\sigma_{T}=4$in. We simulate 10,000 innings for this pitcher following the optimal location policy determined for a pitcher with $\sigma_{F}$. In other words, the pitcher makes decisions under the belief they have control $\sigma_{F}$ but their error is sampled according to $\sigma_{T}=4$in. regardless of the $\sigma_{F}$ value. We investigate the loss of expected runs per inning as a result of the pitcher following various strategies in Figure ~\ref{fig:loss}.

The graph shows that, as expected, a pitcher minimizes their runs allowed by pitching according to a strategy determined by their true control. Interestingly, the loss function is not symmetric around $\sigma_{T}=4$in. We see much higher loss as $\sigma_{F}$ gets smaller, whereas the loss does not grow as fast as $\sigma_{F}$ gets larger. That means that overconfidence in control is more harmful to a pitcher than under-confidence. This is likely because an overconfident pitcher will try to target the corners of the strike-zone without much success, leading to increased walks and increased hits. An under-confident pitcher will target the center of the strike-zone more than they should, leading to more hits but fewer walks.

This analysis led us to try to develop a more accurate and useful control metric. It shows the value of accurately assessing control and being able to communicate that to a pitcher.

\begin{figure}[hbt!]
\centering
\includegraphics[width=0.65\textwidth]{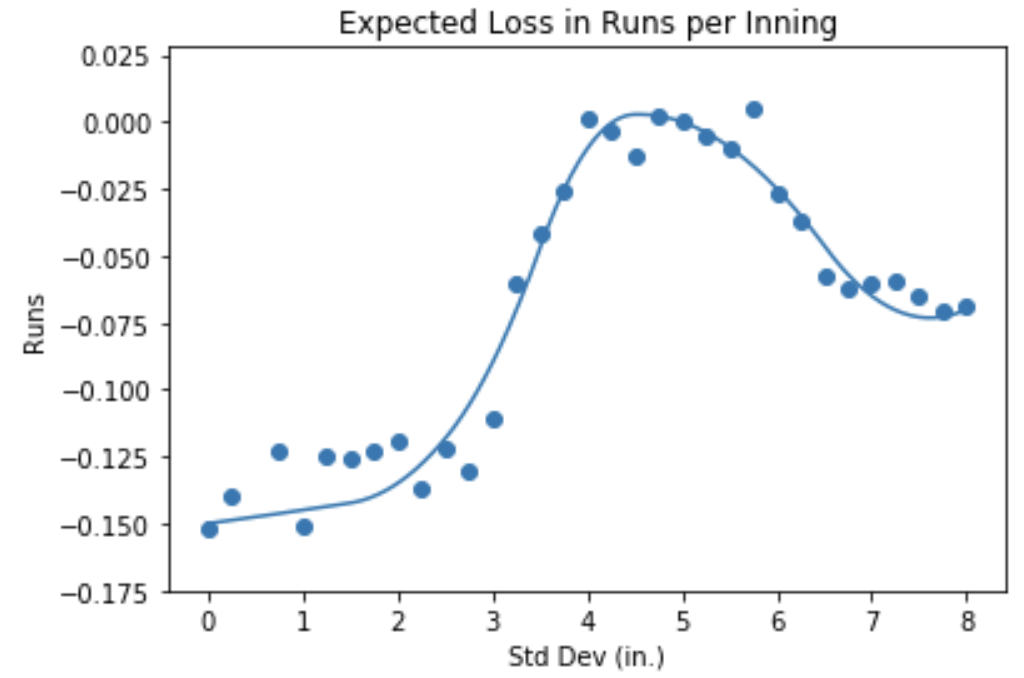}
\caption{
  Loss of expected runs per inning in simulation when a pitcher with $\sigma_{T}=4$ believes he has $x$ control
}
\label{fig:loss}
\end{figure}

\end{document}